
%
\magnification1200


\font\bigbf=cmbx12


\def\degrees{\hbox{${}^\circ$\hskip-3pt .}}
\def\pp{\par\hangindent=.125truein \hangafter=1}
\def\aref#1;#2;#3;#4{\pp #1, {\it #2}, {\bf #3}, #4}
\def\abook#1;#2;#3{\pp #1, {\it #2}, #3}
\def\arep#1;#2;#3{\pp #1, #2, #3}
\def\spose#1{\hbox to 0pt{#1\hss}}
\def\simlt{\mathrel{\spose{\lower 3pt\hbox{$\mathchar"218$}}
     \raise 2.0pt\hbox{$\mathchar"13C$}}}
\def\simgt{\mathrel{\spose{\lower 3pt\hbox{$\mathchar"218$}}
     \raise 2.0pt\hbox{$\mathchar"13E$}}}

\def\frac#1/#2{\leavevmode\kern.1em
 \raise.5ex\hbox{\the\scriptfont0 #1}\kern-.1em
 /\kern-.15em\lower.25ex\hbox{\the\scriptfont0 #2}}

\def\sigsam{\sigma_{\rm sam}}
\def\sigcos{\sigma_{\rm cos}}

\parindent=0pt
\parskip=3pt

\line{\hfil CfPA-TH-93-14}
\line{\hfil astro-ph/9305030}
\line{\hfil 1993 May 24}
\bigskip

\centerline{\bigbf ``Sample Variance'' in Small-Scale CMB Anisotropy
Experiments}
\bigskip

\centerline{Douglas Scott, Mark Srednicki\footnote{*}{On leave
from Department of Physics, University of California, Santa
Barbara, CA 93109} \& Martin White}
\smallskip

\centerline{\it Center for Particle Astrophysics, University of California,}
\centerline{\it Berkeley, CA 94720}

ABSTRACT

We discuss the effects of finite sky coverage and the uncertainty in
extracting information about the power spectrum from experiments on
small angular scales.  In general the cosmic variance is enhanced by
a factor of $4\pi/A$, where $A$ is the solid angle sampled by the experiment.
As a rough guide, an experiment with sensitivity peaking at the
$\ell$th multipole has to cover $\simgt\ell$ independent patches to have a
lower ``sample variance'' than for a whole-sky measurement of the quadrupole.
Our approach gives a relatively simple way of attaching an error bar to the
theoretical prediction for a particular experiment, and thereby comparing
theories with experimental results, without the need for
computationally-intensive Maximum Likelihood or Monte Carlo calculations.

\medskip

{\it Subject headings:} cosmology: cosmic background radiation\ ---\ methods:
statistical

\bigskip
{\bf Introduction}
\medskip

Since the discovery of large angular scale anisotropies in the microwave
background (Smoot et al.~1992) effort has concentrated on constraining the
primordial power spectrum of fluctuations using the results from
degree-scale experiments (e.g.~G{\'o}rski 1992, Vittorio \& Silk 1992,
Gouda \& Sugiyama~1992, Dodelson \& Jubas~1993, G{\'o}rski et al.~1993,
White et al.~1993, Crittenden et al.~1993).
It has become apparent (e.g.~Gould~1993, White et al.~1993) that the
``cosmic variance'', due to sampling only one universe, is a fundamental
limitation on the utility of measurements on the largest scales.
This uncertainty comes about because the fluctuations we observe are a
single realization of a random variable.
The theory can predict properties of the ensemble, but not the individual
realizations.
For example the quadrupole is drawn from a $\chi^2$ distribution with only 5
degrees of freedom, so that the purely theoretical error on the measured
r.m.s.~quadrupole is large.

Usually it is assumed that the ``cosmic variance" is negligible for small
scale ($\simlt1^{\circ}$) experiments which are sensitive only to high
multipoles in the $\Delta T/T$ expansion.
However, this assumption is only true for a whole-sky measurement.  It is
clear that the variance from only sampling a fraction of the sky will be
larger.  We feel that the importance of this ``sample variance'' has not
been emphasized until now.

In this letter we address the issue of the effects of finite sky coverage
for small scale experiments and show that while {\it in principle} the sample
variance can be small, currently it is a major source of uncertainty.
We find that the naive expectation is essentially correct, namely that the
sample variance $\sigsam^2$ is related to the cosmic variance $\sigcos^2$ via
$$\sigsam^2 \simeq (4\pi/A)\sigcos^2\;, \eqno(1) $$
where $A$ is the solid angle covered by the experiment.

Consider an (idealized) experiment which is sensitive only to the
$\ell\,{\rm th}$ multipole.  For a Harrison--Zel'dovich spectrum of
fluctuations, in general we have
$\sigma_{{\rm cos},\ell}/\mu_\ell \sim (\ell + {\frac1/2})^{-1/2}$,
where $\mu_l$ is the theoretical r.m.s. value of the $\ell\,$th multipole.
If this experiment covers $N$ independent patches, each
of size $\sim\ell^{-2}$ steradians, then we have
$${\sigma_{{\rm sam},\ell}\over\mu_\ell} \sim
 \sqrt{{4\pi/A\over\ell + {\frac1/2}}} \sim \sqrt{{\ell/N}}\;.  \eqno(2)$$
Recall that the cosmic variance for a (whole sky) measurement of the
quadrupole gives an uncertainty of order the mean.
Thus, obtaining a measurement with a theoretical
uncertainty below that of the full-sky quadrupole requires $N\simgt\ell$.

Since this rule of thumb is not satisfied for most of the current small
angular-scale experiments, we were motivated to investigate the effect of the
``sample variance'' in more detail.
We find for the MAX (Meinhold et al.~1993, Gundersen et al~1993) and
SP91/ACME (Gaier et al.~1992, Schuster et al.~1993) experiments a theoretical
uncertainty of approximately 25--30\% in the prediction of $\Delta T/T$ on the
appropriate scale.
Many authors use a Bayesian technique (e.g.~Bond et al.~1991), or a Monte-Carlo
simulation (e.g.~G{\'o}rski et al.~1993), in analyzing these experiments.
Such techniques automatically account for the sample variance,
however sometimes the (dominant) source of the uncertainty is obscured.
In the language of Bayesian analysis, our sampling uncertainty indicates
that the data from small patches of the sky are highly correlated (the amount
of experimental ``information" is low), which leads to sensitivity to the
``prior distribution" used in the analysis.

We should stress that this sample variance has nothing to do with the
experimental precision of the measurements, but simply the fact that they
may not necessarily have covered enough of the sky to provide a good
estimate of the r.m.s.~value of $\Delta T/T$ on the relevant angular
scale.  In the rest of this letter we develop this argument more
rigorously, and give estimates for specific experimental configurations.

\bigskip
{\bf Derivation of the sample variance}
\medskip

Consider the correlation function at zero-lag, or
$\left[(\Delta T/T)_{\rm rms}\right]^2$, as estimated by measurements
over a patch of the sky subtending a solid angle $A$ and which we shall
call $\Delta_A$.
We include the effects of finite beam size and possible ``chopping" of the
beam in our definition.

The theory predicts a probability distribution for $\Delta_A$, but not
its value in our universe.
Write $C_0$ as the mean correlation function at zero-lag:
$\langle\Delta_A\rangle=C_0$, where the angled brackets represent
an average over an ensemble of ``universes''.
This is a measure of the amplitude of the
theoretical power spectrum at the scales probed by the experiment.

Under the assumption that the original fluctuations are gaussian,
it is straightforward to calculate
$$\eqalign{
\left\langle\Delta_A^2\right\rangle &= {1\over A^2}\int_A d\Omega_1d\Omega_2
\,\left\langle \tilde{T}(\hat{n}_1)\tilde{T}(\hat{n}_1)
  \tilde{T}(\hat{n}_2)\tilde{T}(\hat{n}_2) \right\rangle \cr
&=  {1\over A^2}\int_A d\Omega_1d\Omega_2 \left[
\left\langle \tilde{T}(\hat{n}_1)\tilde{T}(\hat{n}_1)\right\rangle
\left\langle \tilde{T}(\hat{n}_2)\tilde{T}(\hat{n}_2)\right\rangle
+2\left\langle \tilde{T}(\hat{n}_1)\tilde{T}(\hat{n}_2)\right\rangle^2
\right]. \cr} \eqno(3)
$$
Here $\tilde{T}(\hat{n})$ refers to the temperature difference assigned
to the direction $\hat{n}$ by the experiment, including finite beam width
and possible beam ``chopping".  The angled brackets again represent an
average over an ensemble of ``universes".
To fix the size of the uncertainty in translating from the experimentally
measured $\Delta_A$ to $C_0$ we calculate the sample variance, which is
simply the second moment about the mean for $\Delta_A$:
$$\eqalign{
\sigsam^2 = \Delta^{(2)}_A &= \left\langle\Delta_A^2\right\rangle
                             -\left\langle\Delta_A\right\rangle^2 \cr
                           &= {2\over A^2} \int_{A} d\Omega_1 d\Omega_2
                               \,C^2(\hat{n}_1\cdot\hat{n}_2)\;.\cr} \eqno(4)
$$
where $C(\cos\theta)$ is the 2-point correlation function
$$\eqalign{
 C(\hat{n}_1\cdot\hat{n}_2) &\equiv
 \left\langle \tilde{T}(\hat{n}_1)\tilde{T}(\hat{n}_2) \right\rangle\cr
 &={1\over4\pi}\sum_{\ell=2}^\infty \left\langle a_{\ell}^2\right\rangle
 W_\ell(\hat{n}_1\cdot\hat{n}_2)\;.\cr}\eqno(5)
$$
Here $a_{\ell}^2\equiv\sum_m\left|a_{\ell m}\right|^2$, where the $a_{\ell m}$
are the multipole coefficients of the temperature fluctuation on the sky and
$W_\ell$ is the window function appropriate to the experiment under
consideration
(see Bond et al.~1992, Dodelson \& Jubas 1993, White et al.~1993).

One can also derive the third moment about the mean:
$$
\Delta^{(3)}_A = {8\over A^3} \int_{A} d\Omega_1 d\Omega_2 d\Omega_3
\,C(\hat{n}_1\cdot\hat{n}_2)
\,C(\hat{n}_2\cdot\hat{n}_3)
\,C(\hat{n}_3\cdot\hat{n}_1)\;.\eqno(6)
$$
In general the distribution for $\Delta_A$ will be positively skewed.
Similar expressions can be written for all the higher moments.

\bigskip
{\bf Simple Models}
\medskip

For the purposes of illustration we consider the case of a gaussian
autocorrelation function (GACF).
Specifically we take
$$
C(\cos\theta) = C_0\exp\bigl[(\cos\theta-1)/\theta_c^2\bigr]
              \simeq C_0\exp\bigl[-\theta^2/2\theta_c^2\bigr]\;,\eqno(7)
$$
and we assume $\theta_c\ll 1$.
The correlation function, $C(\cos\theta)$, for both the
SP91/ACME and MAX experiments (to choose specific examples) with CDM-like
power spectra can be adequately approximated by equation (7).
We should emphasize that our $C(\cos\theta)$ has already been convolved with
the telescope beam, thus $\theta_c$ is a (calculable) function of the power
spectrum and the experimental parameters.
Often the {\it unconvolved} $C(\cos\theta)$ is taken to be a GACF, which is
{\it not} a good assumption.

We see from equation~(4) that we expect $\Delta_A^{(2)}\sim C_0^2\theta_c^2/A$
for $A\gg\theta_c^2$.  To explore this further consider a circular region
defined by $0\le\theta<\theta_0$ and $0\le\phi<2\pi$, so that
$A=2\pi(1-\cos\theta_0)$.
Using equations~(4) and (7) we find that
$$
\Delta^{(2)}_A={2C_0^2\over(1-\cos\theta_0)^2}\int_{\cos\theta_0}^1 dc_1 dc_2
\ \exp\bigl[2(c_1c_2-1)/\theta_c^2\bigr]I_0\bigl(2s_1s_2/\theta_c^2\bigr)\;,
                                                           \eqno(8)
$$
where $c_i$ and $s_i$ are $\cos\theta_i$ and $\sin\theta_i$ respectively, and
$I_0$ is the modified Bessel function of order zero.
We have the two limiting cases
$$
\Delta^{(2)}_A \rightarrow \left\{
\matrix{
2C_0^2&{\rm as\phantom{r}}&\theta_0\rightarrow 0\;,\cr
&&\cr
{\displaystyle {1\over 2}\left({4\pi\over A}\right)C_0^2\theta_c^2}
&{\rm for}& \theta_0\gg\theta_c\;.\cr} \right.\eqno(9)
$$
We show $\Delta^{(2)}_A$ and ${1\over 2}(4\pi/A)C_0^2\theta_c^2$ for
$\theta_c=1^{\circ}$ in figure 1.
The expected result is recovered in the limit of one measurement: for a
gaussian random variable $x$, the variance of $x^2$ is twice the square
of the mean of $x^2$.
Notice that above a few degrees the second limiting form is a good
approximation to the exact result.  This means that after a few correlation
lengths the sample variance, $\Delta^{(2)}_A$, approaches
the cosmic variance, ${1\over 2}C_0^2\theta_c^2$, as $4\pi/A$, so that
there is always something to be gained by sampling more sky.

While the region $A$ defined above clearly illustrates the simple scaling of
the sample variance with sky coverage, it is not a good approximation to the
scan strategy of SP91/ACME and MAX.
To consider these experiments more properly we here make a simplified model
of the scan strategies.
In both cases we can approximate the scan pattern as a single line of
(angular) length $\alpha$.  For our GACF assumption we find
$$\eqalign{
\Delta^{(2)}_A &= 2\int_0^{\alpha} {d\phi_1\over\alpha}{d\phi_2\over\alpha}
\,C^2(\cos|\phi_1-\phi_2|) \cr
 &\simeq 2 C_0^2\, {\theta_c\over\alpha}
 \left\{\sqrt{\pi}
 \mathop{\hbox{erf}}\left({\alpha\over\theta_c}\right)
 -{\theta_c\over\alpha}\left[1-\exp(-\alpha^2/\theta_c^2)\right]\right\}. \cr}
 \eqno(10)
$$
For small $\alpha$ we again recover the limit $\Delta^{(2)}=2C_0^2$ while
for $\alpha$ larger than a few $\theta_c$ the second term is small and the
error function is approximately unity.
Comparing with equation~(9) we see that the effective solid angle~$A$
for this linear scan is $\sqrt\pi\theta_c \times \alpha$.
Putting in a MAX correlation angle $\theta_c\simeq{\frac1/2}^{\circ}$ and
a scan length of $\alpha=6^{\circ}$ we find that
$\sigsam=\sqrt{\Delta^{(2)}_A} \simeq 0.5C_0$; that is,
the uncertainty is approximately 50\% of the mean
$\left[(\Delta T/T)_{\rm rms}\right]^2=\langle\Delta_A\rangle=C_0$.
In the case of SP91/ACME, $\theta_c\simeq2^{\circ}$ and
$\alpha\simeq9\times 2\degrees1$, giving a relative error of approximately
60\%.
The actual value of $\theta_c$ depends on the precise cosmological model
assumed, but the values used above are representative for CDM and these
experiments.

The implication of this simple analysis is that the theoretical predictions
for $\Delta T/T$ for the experiments considered should be assigned a relative
error of approximately 25--30\% due to the ``sample variance''.
(However recall that the distribution of $\Delta_A$ will be non-gaussian;
see e.g.~Scaramella \& Vittorio~1991, Cayon et al.~1991, White et al.~1993.)
Other experiments have similarly non-negligible sample variances because of
incomplete sky coverage.  This ``range'' of theoretical predictions should
be considered in comparing theory with small scale experiments.
Only COBE and MIT approach the unavoidable cosmic variance limit for the
scales which they probe.  Note, however, that the galactic cut made by
the COBE team ($|b|>20^{\circ}$) increases the relative error on
$\Delta T/T$ by 23\% over what it would be for a whole sky measurement.

\bigskip
{\bf Conclusions}
\medskip

Our analysis makes clear that the potential gain in theoretical precision on
smaller angular scales is in fact only realized once a significant fraction of
the sky has been covered.
Thus the effects of finite sky coverage are serious for all of the current
small scale experiments: VLA, ATCA, OVRO, MSAM, MAX, SP91/ACME and (on larger
scales) Tenerife.
Including the ``sample variance'' in the theoretical predictions for these
experiments may help to reconcile the apparently contradictory results being
reported on small angular scales (Gaier et al.~1992, Schuster et al.~1993,
Gundersen et al.~1993, Meinhold et al.~1993).
Current analyses of the data (including Monte-Carlo and Bayesian analyses)
correctly take into account the ``sample variance'', but provide little
physical
insight or guide to the size of the effects.  We hope that our approach is
sufficiently straightforward that it will make the comparison of theory and
measurement more transparent for these small scales, and perhaps also act as
an aid for the design of future experiments.

\bigskip
{\bf Acknowledgements}
\medskip

We would like to thank A. Lange, C. Lineweaver, P. Richards and G. Smoot
for useful conversations.
This work was supported in part by grants from the NSF and DOE.

\bigskip
{\bf Figure Caption}
\medskip

Figure: The sample variance, $\Delta^{(2)}_A$, from equation (8) as a
function of sky coverage, $\theta_0$, for a GACF with $\theta_c=1^{\circ}$
(solid line).
Also shown is the naive expectation $(4\pi/A)\sigcos^2$ (dashed line) where
$A=2\pi(1-\cos\theta_0)$ is the solid angle covered by the region sampled
and $\sigcos^2={1\over 2}C_0^2\theta_c^2$ is the cosmic variance.
This is a good approximation to the full result above a few correlation
lengths.

\vfill\eject

\bigskip
{\bf References}
\frenchspacing
\medskip
\aref Bond, J. R., Efstathiou, G., Lubin, P. M. \& Meinhold, P. R., 1991;
Phys. Rev. Lett.;66;2179
\aref Cayon, L., Martinez-Gonzalez, E. \& Sanz, J.L., 1991;MNRAS;253;599
\abook Cheng, E.S., et al.~1993;ApJ;submitted, astro-ph/9305022
\abook Crittenden, R., Bond, J. R., Davis, R. L., Efstathiou, G. \&
Steinhardt, P. J., 1993;Phys. Rev. Lett.;submitted, astro-ph/9303014
\aref Dodelson, S. \& Jubas, J. M., 1993;Phys. Rev. Lett.;70;2224
\aref Gaier, T., Schuster, J., Gundersen, J. O., Koch, T., Meinhold, P. R.,
Seiffert, M. \& Lubin, P. M., 1992;ApJ;398;L1
\aref G{\'o}rski, K. M., 1992;ApJ;398;L5
\abook G{\'o}rski, K. M., 1993;ApJ;in press
\abook G{\'o}rski, K. M., Stompor, R. \& Juszkiewicz, R., 1993;ApJ;submitted,
preprint YITP/U-92-36
\aref Gould, A., 1992;ApJ;403;L51
\aref Gouda, N. \& Sugiyama, N., 1992;ApJ;395;L59
\abook Gundersen, J. O., Clapp, A. C., Devlin, M., Holmes, W., Fischer, M. L.,
Meinhold, P. R., Lange, A. E., Lubin, P. M., Richards, P. L. \& Smoot, G. F.,
1993;ApJ;submitted
\aref Meinhold, P. R., Clapp, A. C., Cottingham, D., Devlin, M.,
Fischer, M. L., Gundersen, J. O., Holmes, W., Lange, A. E., Lubin, P. M.,
Richards, P. L. \& Smoot, G. F.,~1993;ApJ;409;L1
\aref Readhead, A. C. S., Lawrence, C. R., Myers, S. T., Sargent, W. L. W.,
Hardebeck, H. E., \& Moffet, A.T., 1989;ApJ;346;566
\aref Scaramella, R. \& Vittorio, N., 1991;ApJ;375;439
\abook Schuster, J., Gaier, T., Gundersen, J. O., Meinhold, P. R., Koch, T.,
Seiffert, M., Wuensche, C. A. \& Lubin, P. M., 1993;ApJ;submitted,
preprint UCSB
\aref Smoot, G., et al., 1992;ApJ;396;L1
\aref Vittorio, N. \& Silk, J., 1992;ApJ;385;L9
\abook White, M., Krauss, L. \& Silk, J., 1993;ApJ;submitted,
astro-ph/9303009

\bye